\documentclass[a4paper,11pt]{article}
\usepackage{pos}

\title{Formal Theory: Status and Outlook}
\author*[a]{Fernando Quevedo}
\affiliation[a]{Department of Applied Mathematics and Theoretical Physics,\\
  University of Cambridge, Wilberforce Road, Cambridge, CB3 0WA, UK}

\emailAdd{f.quevedo@damtp.cam.ac.uk}
\abstract{A very brief overview is presented on some of the current most active areas of research in formal aspects of High Energy Physics and Cosmology. Including the recent 
breakthrough on the black hole information paradox, developments on amplitudes, the bootstrap and swampland programmes as well as progress towards realistic UV complete models of particle physics and cosmology. Perspectives on future contact with observations are discussed emphasising the long term prospect for ultra high frequency gravitational waves to test early universe physics beyond the Standard Model.}

\FullConference{%
  41st International Conference on High Energy physics - ICHEP2022\\
  6-13 July, 2022\\
  Bologna, Italy
}

\begin{document}
\maketitle

\section{Introduction}
I would like to congratulate  the organisers of ICHEP-2022 for managing to put together and implement such an ambitious  post-pandemic programme. Let me start by paying tribute to arguably the most
 prominent figure in our field, who unfortunately left us after the previous ICHEP conference: Steven Weinberg. It has been  my greatest honour to have him as my PhD supervisor. Preparing for this talk I went to read his two summaries of previous ICHEPs (1986 \cite{Weinberg:1986nj} and 1992 \cite{Weinberg:1992dq}). %when ICHEP was also referred to as the Rochester conference. 
 There, he  claims that this is actually a very easy task because everybody knows it is impossible to summarise such a big conference and people even forgive you if you do not refer to them. %Even after passing away I keep learning from Steve. As usual, both summaries, even though a bit outdated, are a pleasure to read and give us an idea of the open questions at that time and his general vision for the field.

 Here I hope to convey the main directions in the subfield roughly defined by the {\it hep-th arxives}, by briefly summarising the main recent achievements, providing some relevant references (mostly reviews) for the interested reader, which I assume to be a high energy experimentalist or phenomenologist. A way to roughly measure the progress and change of perspective over the past few years is to compare this talk with   that given by Joe Polchinski \cite{Polchinski:2008ux} 14 years ago, which follows  similar lines. %as I follow here. 
 Broadly, we may classify the subjects in two groups, as reflected by  the different parallel session talks at this conference:
  {\it Formal aspects of  Quantum Field Theories (QFT)} and 
 {\it Quantum Aspects of Gravity}.  
 %\end{enumerate}
  Recent progress is summarised in \cite{Craig:2022cef,Bah:2022wot} and \cite{deBoer:2022zka} respectively.

 %\vspace{-0.2 cm}

  \section{Holography, Black Holes and Information}
  
The most important achievement in theoretical physics during  the past 25 years has arguably  been the  AdS/CFT duality, as proposed by Maldacena in 1997. %\cite{Maldacena:1997re}.
 The idea refers to the equivalence between two apparently different theories: the bulk gravitational theory in $d+1$ dimensions and the non-gravitational boundary theory in $d$ dimensions. The standard analogy for this duality is with a can of soup for which all the information about the soup (bulk) is encoded in the writings on the cover of the can (the boundary). The typical example being the non-gravitational conformal field theory corresponding to ${\mathcal{N}}=4$ supersymmetric Yang-Mills theory in four-dimensions  dual to  string theory on a product of a five dimensional anti de Sitter space $AdS_5$ and a five-dimensional  sphere $S_5$  (to complete the ten dimensions of string theory). This provides us a concrete example of a consistent quantum theory of gravity through its equivalence with the well-defined non-gravitational field theory. Furthermore, it gives an explicit realisation of the general expectation that, in a full theory of quantum gravity, spacetime could be an {\it emergent} entity. In this case the extra spatial dimension of the bulk emerges with respect to the boundary. Hence the term holography.

%\vspace{-0.5cm}

 \begin{figure}[ht]
   % \centering
    \includegraphics[width = 0.4\textwidth]{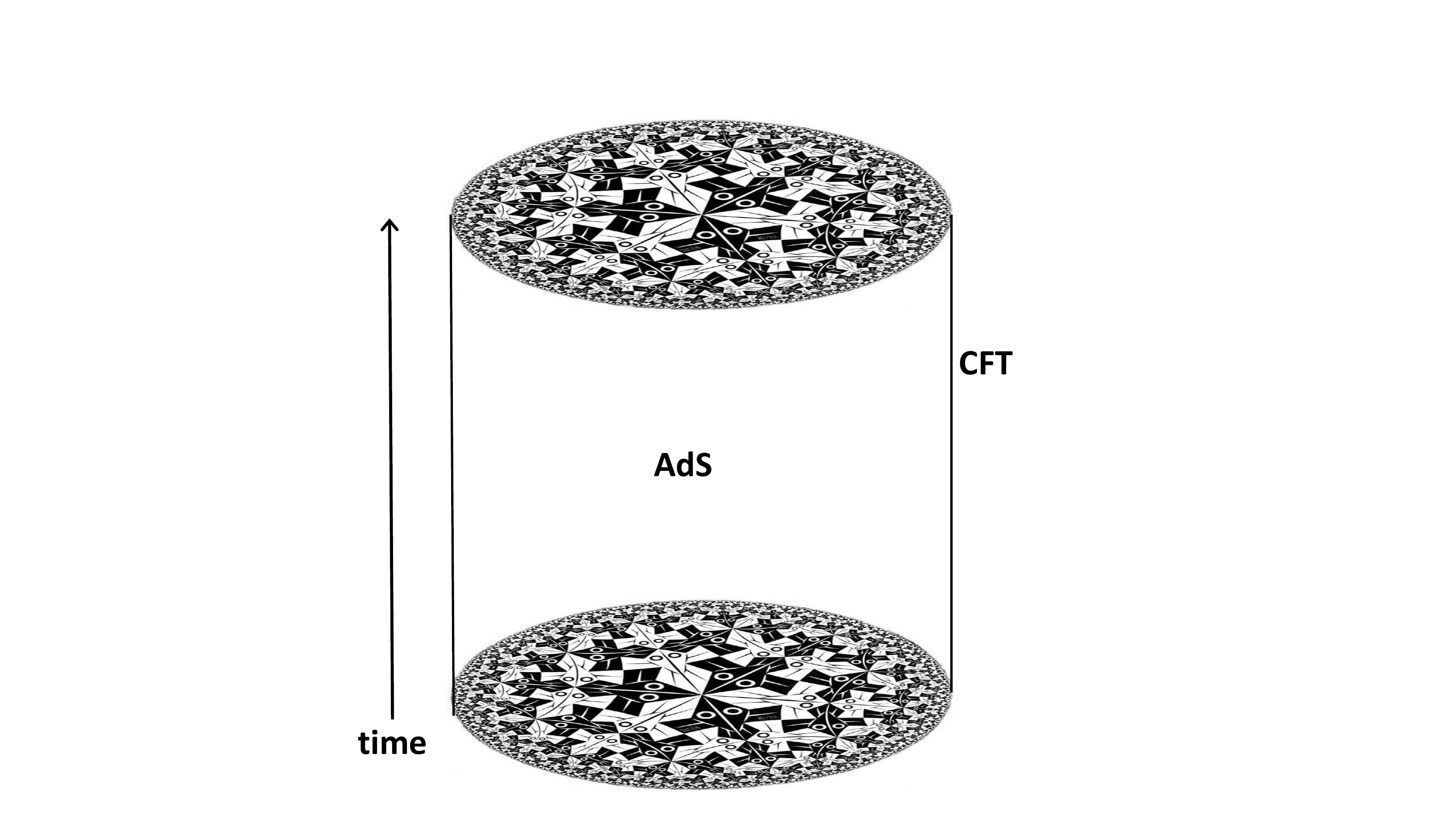} 
     \includegraphics[width = 0.2\textwidth]{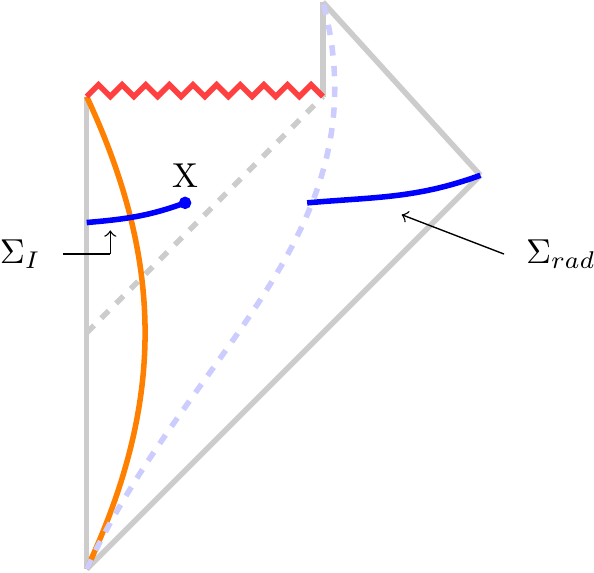}
      \includegraphics[width = 0.6\textwidth]{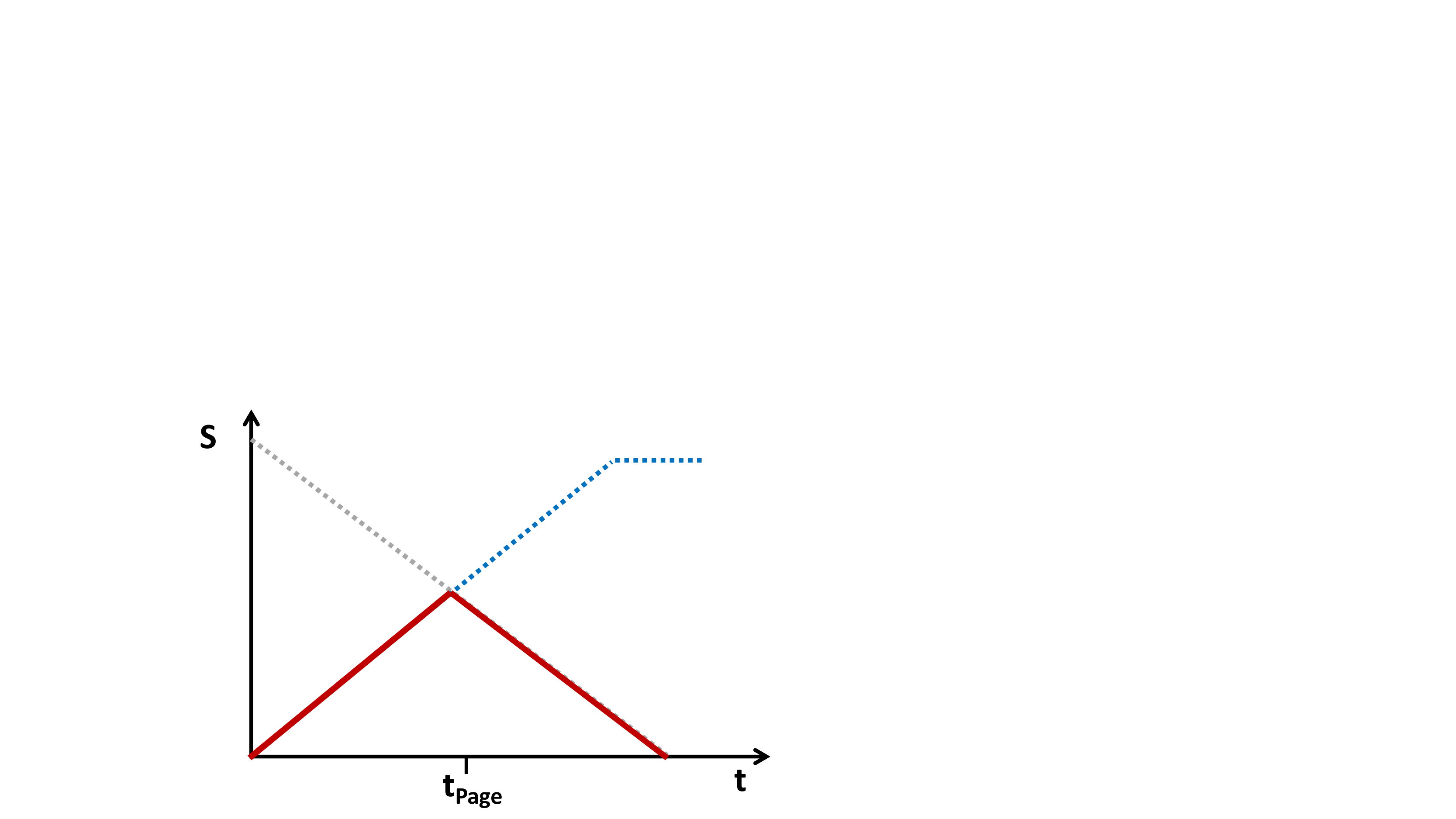}
    \caption{\footnotesize{Left: A pictorial representation of the AdS/CFT duality. Center: A pictorial representation of  recent progress   on black hole information. A region inside the horizon, known as an Island
    $\Sigma_I$ bounded by a quantum extremal surface (QES) X, stores the degrees of freedom that  keep the information that has not reached the Hawking radiation (figure made by Veronica Pasquarella). Right: The time evolution of the entropy (in red) starting to increase but then decrease at the Page time when the island configuration provides the extremum of the entropy and hinting at a unitary evolution %(an  initial pure state ($S=0$) to a final pure state)
    .}}
    \label{fig:AdSCFT}
\end{figure}

\subsection{Islands, Page curve and no loss of information}

Probably the most important application of AdS/CFT duality regards the Hawking black hole information paradox (for recent reviews see for instance \cite{Almheiri:2020cfm,Raju:2020smc,Bousso:2022ntt,Harlow:2022qsq}). Let us state what the original problem is. The black hole entropy, as determined by Bekenstein and Hawking, is given by the area formula:
$
S_{BH}=\frac{kc^3}{4G\hbar}A 
$
with $c, k, \hbar, G$ the  speed of light, and the Boltzmann, Planck  and gravitational constants respectively and $A$ the area of the black hole  horizon. Note that this expression already encodes the idea of holography since the entropy, which counts the number of states in a given volume, is determined not by the volume but by the area.

 Hawking radiation with energy density $\rho$ has entropy  given by the von Neumann entropy:
$
S_R=\rho\log\rho \leq S_{BH}.
$
 The information loss paradox amounts to the fact that, while the black hole evaporates, $S_R$ increases monotonically. If  the system  starts as a pure state with zero entropy and information were preserved,  it should end with a zero entropy state instead of increasing. %\footnote{For a recent non-technical discussion on successful microscopic counting of black hole states from AdS/CFT see \cite{PandoZayas:2020iin}.}. Furthermore, while the black hole evaporates, its area decreases $A\rightarrow 0$ making the general property $S_R\leq S_{BH}$ impossible to satisfy. 
 %This is  the Hawking information loss paradox. 

If holography holds  then there should not be  information loss in black hole evaporation, simply because the boundary  theory is clearly unitary since it is an ordinary field theory.  This argument essentially turned the debate about information loss to rest at the turn of the milenium. However there is  a need to understand how information is preserved directly from the black hole perspective.

The progress in the past 3 years refers to use a generalised expression for the radiation entropy in terms of the area of what is called a quantum extremal surface (QES). 
$
S_{\rm gen}=\frac{kc^3}{4G\hbar}A(X)+S_{\rm bulk}
$
where  $X$ is the QES and $ S_{\rm bulk} $ is the von Neumann entropy of a region bounded by $X$. Typically this expression does not give a unique expression for $S_{\rm gen}$ and the prescription is to take the extremal value (hence the QES name). While the black hole evaporates, this entropy increases monotonically, with the contribution of $ S_{\rm bulk} $ being originally very small, however at a critical time, known as {\it Page time}, a different QES appears to extremise the entropy in such a way that the QES is the boundary of a region inside the black hole horizon known as an {\it island}. This region, denoted as $\Sigma_I$ in the figure above, is clearly disconnected from the region outside the horizon where the Hawking  radiation is emitted. 
 The relevance of this new solution is that the corresponding entropy starts increasing as in the Hawking calculation but it decreases after the Page time and reduces to zero, consistent with unitarity. This is a major step towards explaining the no loss of information in black holes. It has also lead to the first attempts towards exploring the potential impact that island-like configurations could have in cosmology (see for instance \cite{Hartman:2020khs}).

\subsection{Generalised Holography}

Following the developments on black hole information, in recent years several generalisations of AdS/CFT duality have been proposed which open up a much broader meaning of holography and emerging spacetime 
with potentially interesting applications:
\begin{itemize}

\item{\it dS/CFT and dS/dS duality.} It is natural to ask if there exists a duality for de Sitter (dS) spacetimes that resemble our current universe. %and also an early inflationary epoque. 
A first concrete proposal for dS/CFT duality was made in \cite{Strominger:2001pn} but contrary to the AdS case there is no consensus about the nature of this duality %In particular the boundary could be a spacelike surface at infinity or even the de Sitter horizon as recently proposed in \cite{Susskind:2021omt}. Furthermore, unlike the AdS case %in which there are concrete examples from string theory solutions,
 and contrary to AdS/CFT there are no concrete realisations of dS/CFT. A  related  proposal is the dS$_{d+1}$/dS$_d$ duality \cite{Alishahiha:2004md} for which the boundary theory is another dS. %in one dimension less. %in which the boundary theory is another de Sitter spacetime in one less dimensions. Recent progress in this proposal was reported in \cite{Gorbenko:2018oov}.

\item{\it Minkowski and Celestial Sphere duality.}
We may also wonder  if there is a duality associated to flat spacetime rather than AdS. This has attracted much attention recently after the proposal of {\it Celestial holography} for which the boundary theory is the celestial sphere at infinity
(for a review see \cite{Pasterski:2021raf}). 
%This proposal was based in part by the observation that a rich class of symmetries are present for asymptotically flat spacetimes and conservation laws for these large gauge symmetries are equivalent to Weinberg's soft theorems %(that explain charge conservation, the equivalence principle and the absence of long range interactions mediated by helicities greater than 2 in QFT). The proposal that gravity in asymptotically flat spacetimes is dual to a theory in the celestial sphere at infinity. 
Again, unlike AdS/CFT there is no concrete realisation of this duality (see however  \cite{Costello:2022jpg}), but contrary to the dS case, we understand better direct calculations in flat spacetimes.
% and this duality has led to interesting insights regarding the S-matrix and amplitudes.

\item{\it $T\bar T$ deformations and duality.}
A further generalisation of AdS/CFT duality is to consider the duality not for the whole AdS spacetime but for a portion of it. The corresponding dual may  not be a CFT but a deformation of a CFT known as $T\bar T$ deformation \cite{McGough:2016lol}. %that has recently been developed within the CFT community.

\item {\it Wedge holography.} A proposal for a wedge spacetime in $d+1$ dimensions to be dual to a quantum theory in $d-1$ dimensions located at the corner of the wedge indicating a co-dimension 2 duality \cite{Akal:2020wfl}. This bulk reconstruction from a co-dimension 2 spacetime has played an important role in the understanding of black holes information.

\end{itemize}

Furthermore, recent developments on von Neuman algebras and holography are opening new  ways to understand the mysteries behind AdS/CFT and QES (see for instance \cite{Chandrasekaran:2022eqq} for a recent discussion).

\section{Towards UV complete models of particle physics and cosmology}

We all know that the Standard Model (SM) of particle physics  is only valid as an effective field theory (EFT) at energies below the Planck scale $M_p=\sqrt{\frac{\hbar c}{G}}\simeq 10^{19}$ GeV. Deriving the SM or its possible realistic extensions  from a UV complete theory has been a major challenge, especially since the mid 1980's when string theory became the  concrete candidate for a unified theory of all particles and interactions. A fundamental theory should address the main questions regarding the SM such as explaining:  
%\vspace{1cm}
\begin{center}
%{\bf Challenges for UV complete theories}\vspace{0.3 cm}
\begin{tabular}{ | l | l | l }
 Standard Model gauge and matter spectrum & Hierarchy of masses & Dark energy \\ 
  Inflation or alternatives & Hierarchy of couplings & Dark matter \\  
  Baryogenesis and proton decay & Flavour: CKM, PMNS  matrices & Dark radiation  
\end{tabular}
\end{center}

During more than 30 years much effort has been devoted to these challenges within string theory with partial success. See for instance the comprehensive textbook \cite{Ibanez:2012zz}.

\subsection{Moduli stabilisation}
String theory adds to the challenges mentioned above at least one extra major challenge, namely,  {\it moduli stabilisation}. This refers to the problem to determine dynamically the size and shape of the extra dimensions predicted by string theory.
%fact that the dimensionality of spacetime is not put in by hand (as we were used to in most theories) but it is a prediction of the theory that the universe is of 10 or 11 dimensions and the extra 6 or 7 dimensions need to be stabilised at small values to be consistent with the fact that at distances larger than those explored at the LHC ($\simeq 10^{-16}$ cms we only see 4 spacetime dimensions. Solving this challenge was the main stumbling block for any theory formulated in extra dimensions. A major progress was made in the early years of this century by considering flux compactifications for which quantised fluxes similar to magnetic fluxes in electromagnetism manage to stabilise the extra dimensions that they are threading. These combined with the consideration of quantum (perturbative and non-perturbative) to the low energy string  EFT can fix the hundreds or thousands of moduli fields corresponding to the shape and size of the extra dimensions. In some cases their dynamical determination give rise to universes similar to  ours with only 4 observable extra dimensions.
In principle, none of the challenges mentioned above can be addressed without solving the moduli stabilisation problem. The breakthrough in this direction came in the first decade of this century in terms of flux compactifications (see \cite{Douglas:2006es} for a review). These compactifications can fix the moduli but generically at a negative value of the vacuum energy. Extra ingredients, such as the introduction of anti branes (see figure) have been considered in order to get positive vacuum energies. This gives rise to de Sitter space and the  string landscape that, due to the huge number of flux compactifications,
can address the dark energy problem. We may say that this is  {\it the worst solution of the dark energy problem with the exception of all the others,} since, so far, it is the only proposal that addresses the main part of the problem which is the contribution to the vacuum energy from each of the Standard Model fields. This is one of the most active research areas in this field given the importance of the problem (dark energy) and the fact that the working scenarios are necessarily contrived and rely on approximations that are well justified but not under full control, so much work remains to be done before claiming full success.
\begin{figure}[ht]
   % \centering
      %\includegraphics[width = 0.6\textwidth]{Swampland.pdf} 
      \includegraphics[width = 0.55\textwidth]{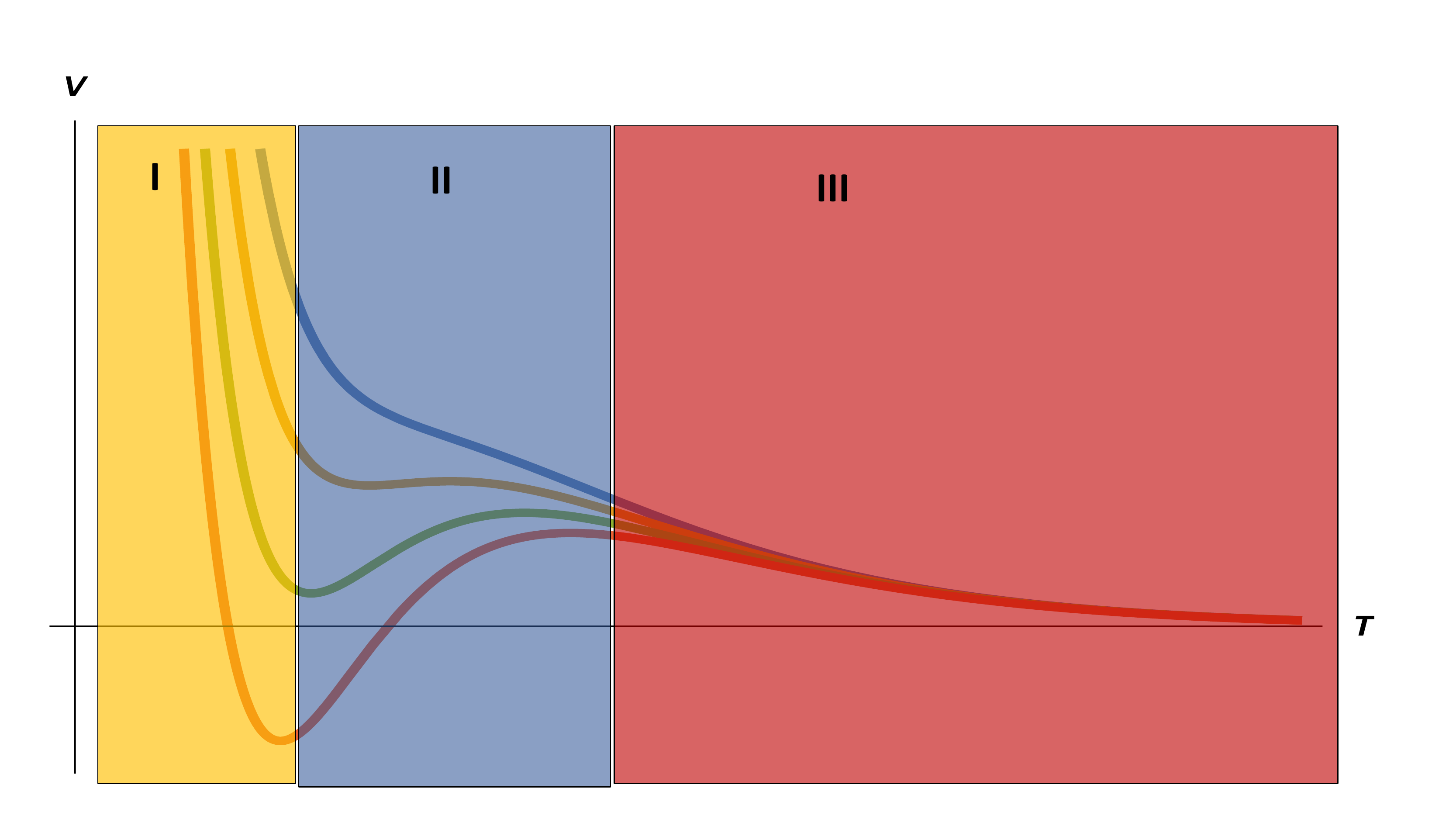}\qquad\qquad\qquad
       \includegraphics[width = 0.3\textwidth]{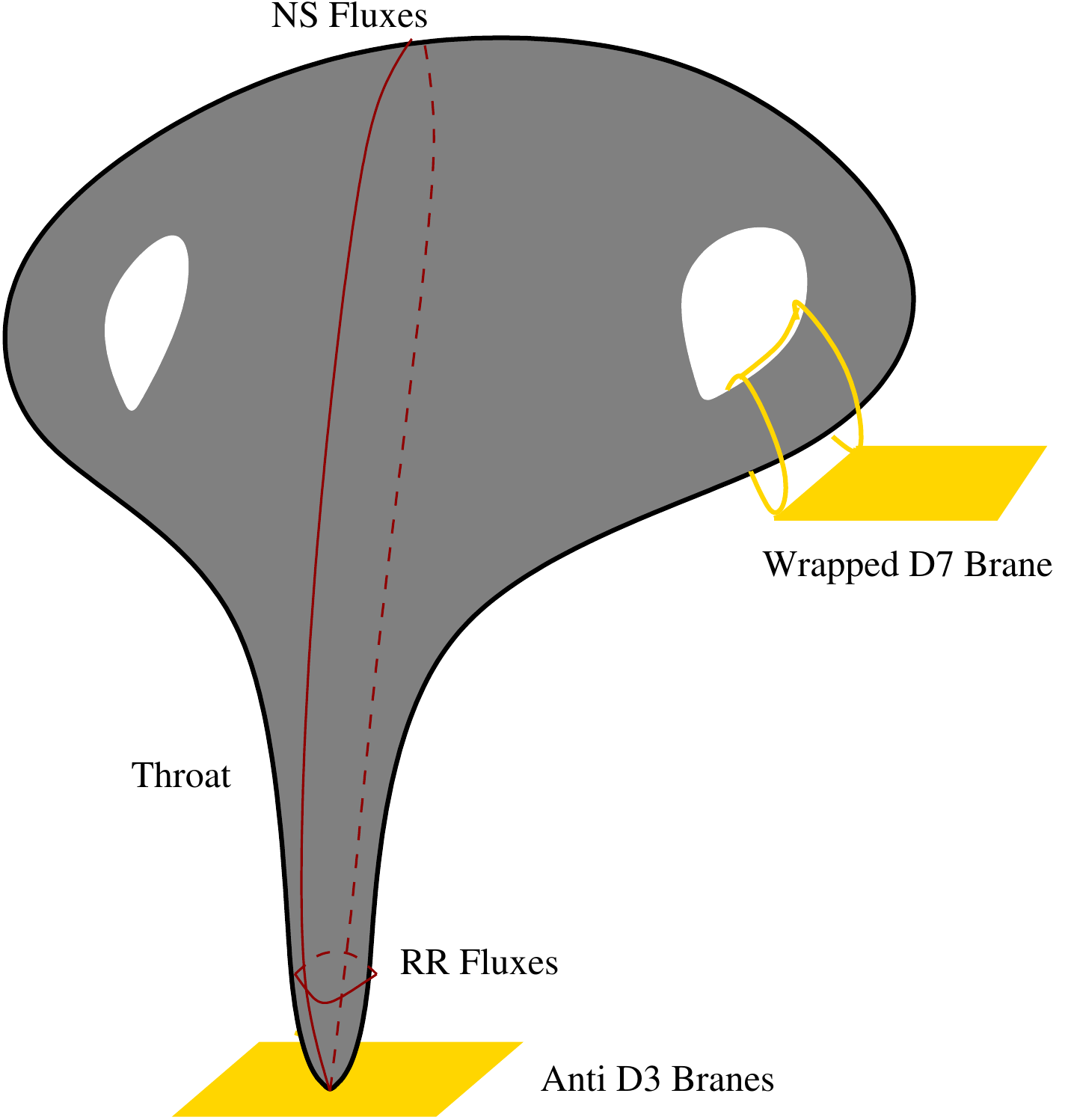}
    \caption{\footnotesize{Left: The Dine-Seiberg problem for moduli stabilisation:  the runaway region III is the only one where EFT is fully trusted. The physically interesting region II requires compactification inputs, like fluxes, to achieve a minimum. Region I would correspond to strong coupling/small volume out of reach from EFT. Right: A typical configuration of fluxes and (anti) branes as in the standard KKLT and LVS scenarios (taken from \cite{Cascales:2003wn}).}}
    \label{fig:moduli}
\end{figure}

\subsection{Recent progress on particle physics models}
These are  examples of  recent progress on particle physics models from string theory. 
\begin{itemize}
\item{\it F-theory models.} F-theory techniques have been developed to allow the construction of a huge number of models ($N=10^{15}$) with the spectrum of the MSSM  \cite{Cvetic:2019gnh}. This is clear progress towards a fully realistic string theory model. But  these models do not include moduli stabilisation and this remains the open question.

\item{\it Heterotic models.} Heterotic string models with MSSM spectrum have also been  obtained with a number as big as $N>10^{23}$ \cite{Constantin:2018xkj}. Again the lack of moduli stabilisation prevents to claim full success.

\item{\it IIB string models.}  Concrete string models including the MSSM and left-right symmetric generalisation were recently obtained in \cite{Cicoli:2021dhg} including moduli stabilisation. The couplings are small enough to partially trust the perturbative expansion. But further generalisations and determination of couplings (such as Yukawa couplings) have not been studied yet.

\item{\it Eclectic Flavour models.} Study of flavour and modular symmetries in heterotic models is allowing to uncover potential patterns to address the questions regarding the difference between different flavours of quarks and leptons. This is a nice example of mutual feeding between pure phenomenology and string theory \cite{Baur:2022hma}.

\item{\it Computational techniques and string models.} A very recent development includes the use of machine learning, genetic algorithms and other computational techniques to study string models. A concrete recent achievement is the explicit calculations of the metric of Calabi Yau spaces, a challenge that has been opened for many years (for a review see \cite{Ruehle:2020jrk}). Furthermore,  obtaining exponentially small superpotentials  in a systematic way was achieved at \cite{Demirtas:2021nlu}.
\end{itemize}

%\subsection{Naturalness and Dark Energy}

\subsection{String Inflation and post-inflation}

%\begin{figure}[ht]
   % \centering
   %  \includegraphics[width = 0.6\textwidth]{Inflation.pdf}
      %\includegraphics[width = 0.5\textwidth]{moduli_domination3.pdf} 
    %  \includegraphics[width = 0.6\textwidth]{Page.pdf}
    %\caption{\footnotesize{Left: A typical inflationary scenario with $\phi$ as the inflaton field being an open or closed string modulus. Right: Moduli domination after inflation. Contrary to standard cosmology instead of reheating from inflation, the reheating of the universe comes from the last of the moduli fields $\varphi$  to decay. Generically $\varphi\neq \phi$. Leading to a period of matter domination after inflation and before reheating.}}
    %\label{fig:ModuliDomination}
%\end{figure}

Inflation is an extremely successful ad-hoc scenario for early universe but it needs to be derived from  a fundamental framework. In string theory, inflation can only be addressed after moduli stabilisation is achieved. Several proposals for string inflation have been  put forward. A summary of the main models is presented in the table. %, with the existing dozen or so string scenarios of inflation with their predicted values of the spectral index $n_s$ and tensor to scalar ratio $r$ extracted from \cite{Burgess:2013sla}. We also show the most recent experimental bounds that are able to essentially rule out some of the string scenarios. 
%In particular, axion monodromy is very much constrained by the value of $r<0.036$. Fibre inflation is the next to be tested. Other models such as racetrack inflation are essentially ruled out by the improved bounds on $n_s$. 
 Next generation of CMB experiments will be able to fully test the two models with large $r$ in the list, namely axion monodromy and fibre inflation.

\vspace{-0.4cm}

\begin{figure}[ht]
   % \centering
      %\includegraphics[width = 0.6\textwidth]{Swampland.pdf} 
      \includegraphics[width = 0.65\textwidth]{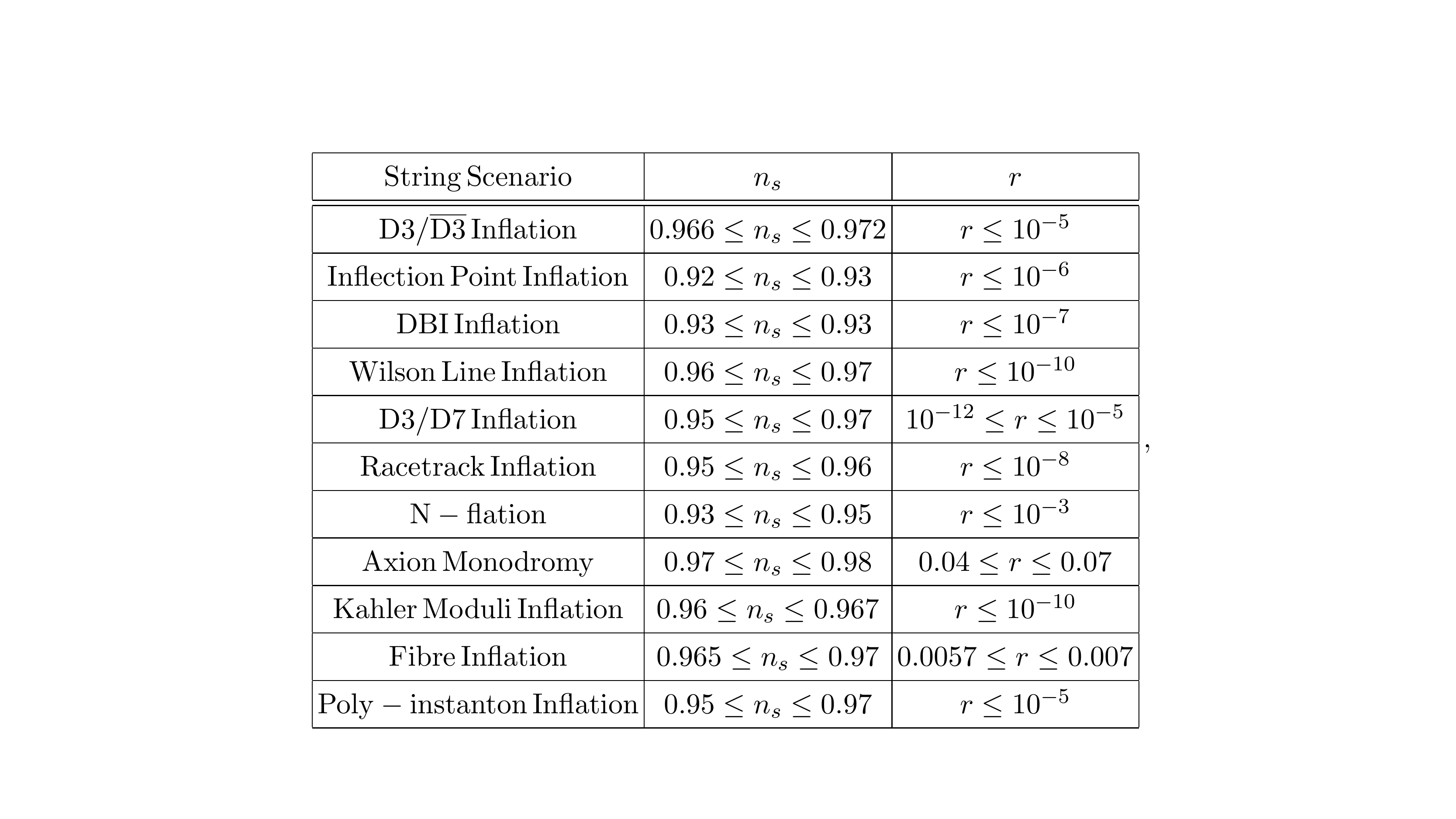}
       \includegraphics[width = 0.7\textwidth]{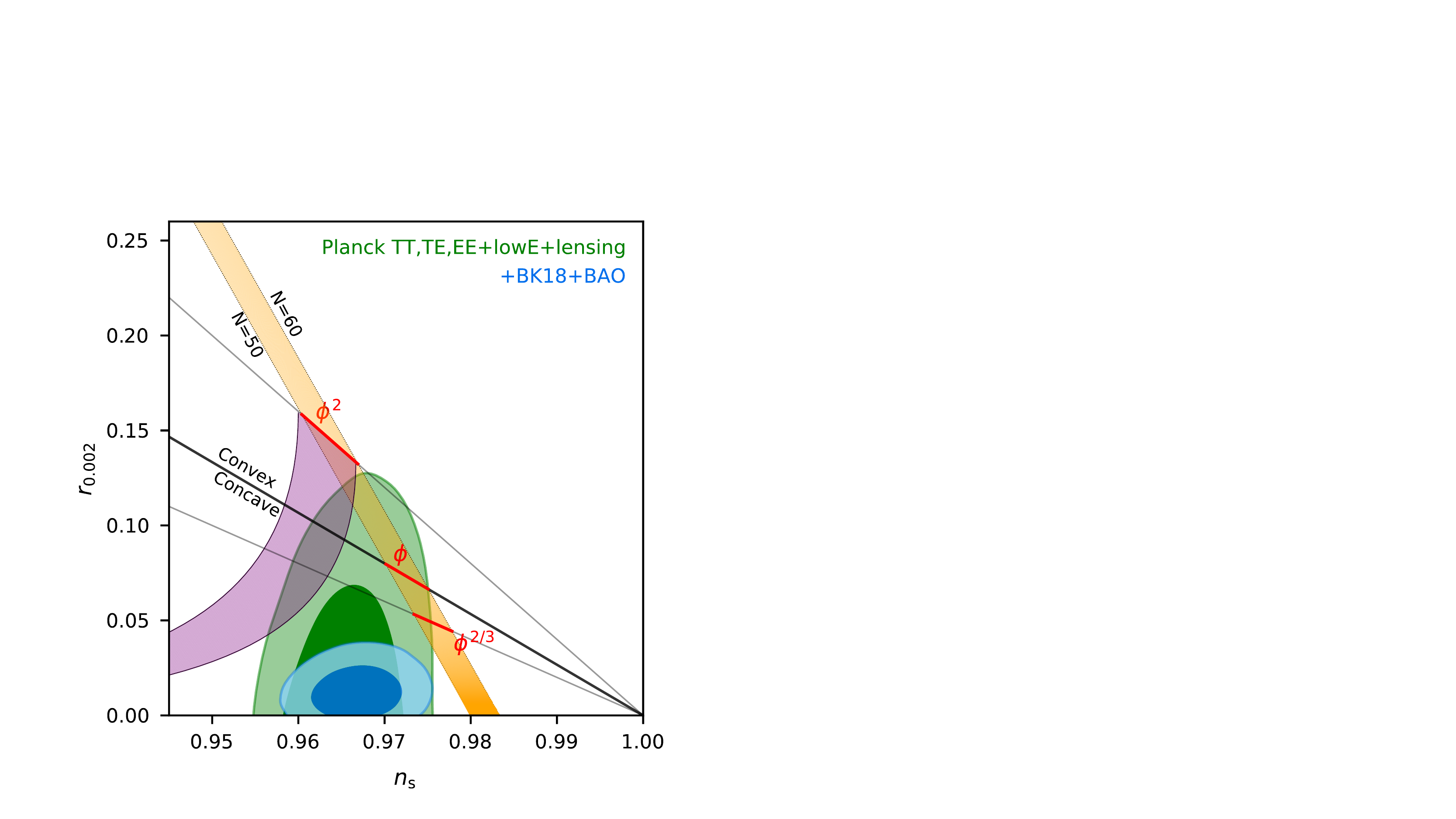}
    \caption{\footnotesize{Left: A table of the stringy inflationary scenarios that have been proposed together with their generic predictions for the two observables: the spectral index $n_s$ and the tensor to scalar ration $r$. Taken from \cite{Burgess:2013sla}. Right: the latest measurements from BICEP/PLANCK \cite{BICEP:2021xfz} which substantially restricts the allowed values of $r$ and $n_s$. Several models can be seen to be in tension with the experiments.}}
    \label{fig:CMB}
\end{figure}

%\subsection{Post-Inflation and Gravitational Waves}
A property of inflation is that the exponential expansion tends to dilute all the physics at scales higher than the inflation scale. This is good since we do not need to know the full string theory to discuss inflation in terms of an EFT, but it is also bad, since inflation may hide all the underlying string physics. Fortunately, the moduli fields tend to be light and survive after inflation leaving a string imprint at low energies. They are the best candidates to be the inflaton but may also play an important role after inflation (for a forthcoming  review see \cite{review}).% Furthermore, besides being candidates to be the inflaton, if the lightest modulus is not the inflaton, it will dominate the energy density after inflation since it will start oscillating about its true minimum. This may have interesting implications differing from standard cosmology (for a recent review see \cite{review}).

The main post-inflation implications of moduli fields include: moduli domination of the energy density ($\rho\sim 1/a^3$)  after inflation; a potential period of {\it kination} with kinetic energy of the moduli ($\rho\sim 1/a^6$) dominating; production of dark radiation after moduli decay and the production of inhomogeneities while the moduli oscillate around their minima, known as {\it oscillons} or {\it oscillatons} (or {\it boson stars}) if gravity is responsible for their quasi-stability. These inhomogeneities can give rise to gravitational waves of frequencies in the Mega to Giga Hertz region.

The high frequencies found in oscillons spectrum motivated  a systematic approach, independent of string theory,  to ultra  high frequency gravitational waves UHF-GW (recall Earth interferometers explore frequencies below the kHz and LISA will explore even lower frequencies)  as a way to search for early universe physics beyond the SM \cite{Aggarwal:2020olq}. This includes, besides oscillons, phase transitions, cosmic strings, preheating, primordial black holes, etc. This, in the long future, may be our main way to test  theories at energies as high  as needed to test  quantum theories of gravity. The long term nature of this initiative may be challenging but prospects look better than LIGO looked 40 years ago and certainly better than any collider initiative. It is then worth pursuing creative ways of detecting high frequency gravitational waves, a field that has been largely unexplored.

%\vspace{-0.7cm}

\begin{figure}[ht]
    \centering
     \includegraphics[width = 0.99\textwidth]{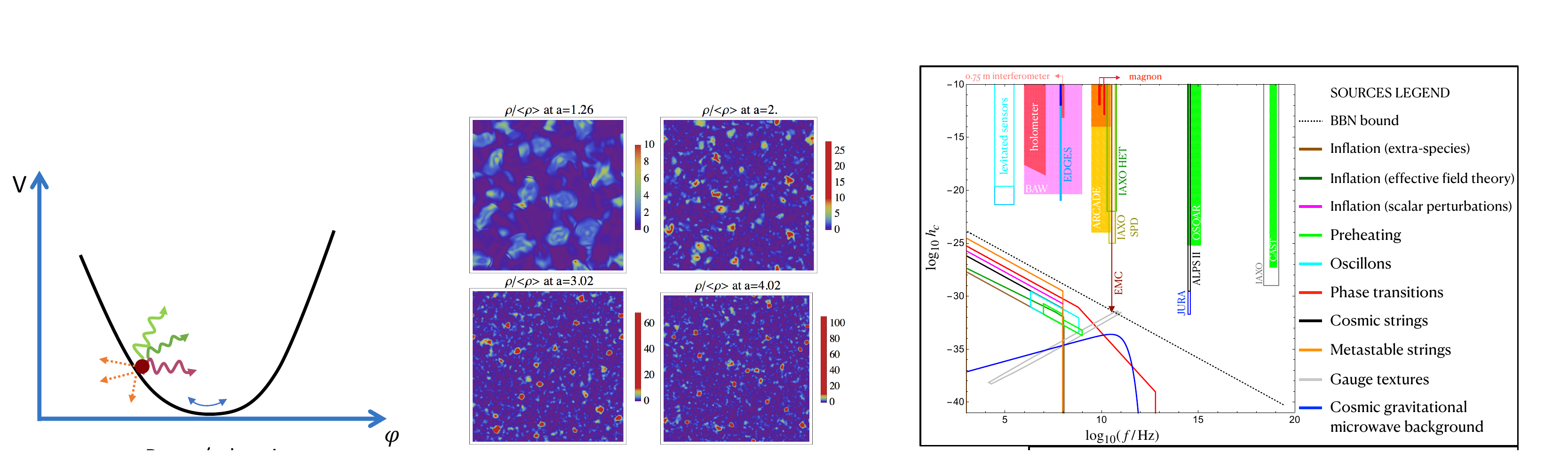}
    \caption{\footnotesize{Left: Moduli domination. After inflation one modulus field starts oscillating and dominate the energy density of the universe. Center: Example of oscillons produced by moduli fields while oscillating around their minima taken from \cite{Antusch:2017flz}. 
    The asymmetry in the density of the configuration implies a production of gravitational waves at high frequencies in the order of the GHz. Right: A general study of challenges and opportunities for stochastic high frequency gravitational waves with different potential sources \cite{Aggarwal:2020olq}.}}
    \label{fig:HFGW}
\end{figure}

%\vspace{-0.1cm}

\section{Amplitudes/Bootstrap and Swampland}

Let us finally say a few words of three of the most active research areas at present: amplitudes, the bootstrap and swampland programmes. It is impossible to properly summarise them in this short space and we will content ourselves by indicating the recent reviews that have appeared on these fields for further reading.
 
\begin{enumerate}
\item{\bf Amplitudes.} Amplitudes are the bread and butter of particle physics. In the past few years a compact and energetic community of theorists has organised itself into a major project aiming at developing tools towards computing S-matrix amplitudes. This is needed for experiments such as LHC, for gravitational waves and for more formal aspects such as supersymmetric theories and the deep programme towards what is known as the amplitudehedron which may be a way to untangled the fundamental theory of nature. For a recent review see \cite{Bern:2022jnl}. There exists also a very active community organised through a collaboration known as SAGEX:https://sagex.org/.

\item{\bf Bootstrap.}
%\subsection{Amplitudes/Bootstrap}
One aspect of the amplitudes initiative is the bootstrap programme in which general requirements such as causality, analiticity and unitarity impose strong constraints on general theories as it was popular before the advent of QCD but now with broader perspectives on field theory and gravity and with stronger computational tools in order to constraint the valid EFTs. For a recent review see:\cite{Kruczenski:2022lot}. This programme has been extended to cosmology in terms of the cosmological bootstrap which is one of the most active areas of theoretical cosmology, extracting model independent properties of inflation and de Sitter space, which is only starting to be developed \cite{Baumann:2022jpr}.

\item{\bf Swampland.} Finally, one of the most active areas in the past few years has been the swampland programme in which a difference is made between consistent EFTs that can be UV completed (the landscape) and those that cannot (the swampland). The programme consists of a series of conjectures which are educated guesses based on generalisations of known properties of string theory that could be promoted to general properties of quantum gravity theories. This is a broad area, for the most recent review see \cite{vanBeest:2021lhn}. The main conjectures include:
(i) {\it Gravity as the weakest force;}   %\cite{Harlow:2022gzl}.
(ii) {\it No global symmetries;}
(iii) {\it Distance conjecture;} 
(iv) {\it de Sitter conjecture;} 
(v) {\it Trans-Planckian conjecture;} 
(vi) {\it Cobordism conjecture.} 
We refer the reader to the review \cite{vanBeest:2021lhn} and references therein for further details on these conjectures. 
%which illustrate an interesting approach towards searching for relations of quantum gravitational theories (not necessarily string theory) to the observable universe.

\begin{figure}[ht]
    \centering
      \includegraphics[width = 0.4\textwidth]{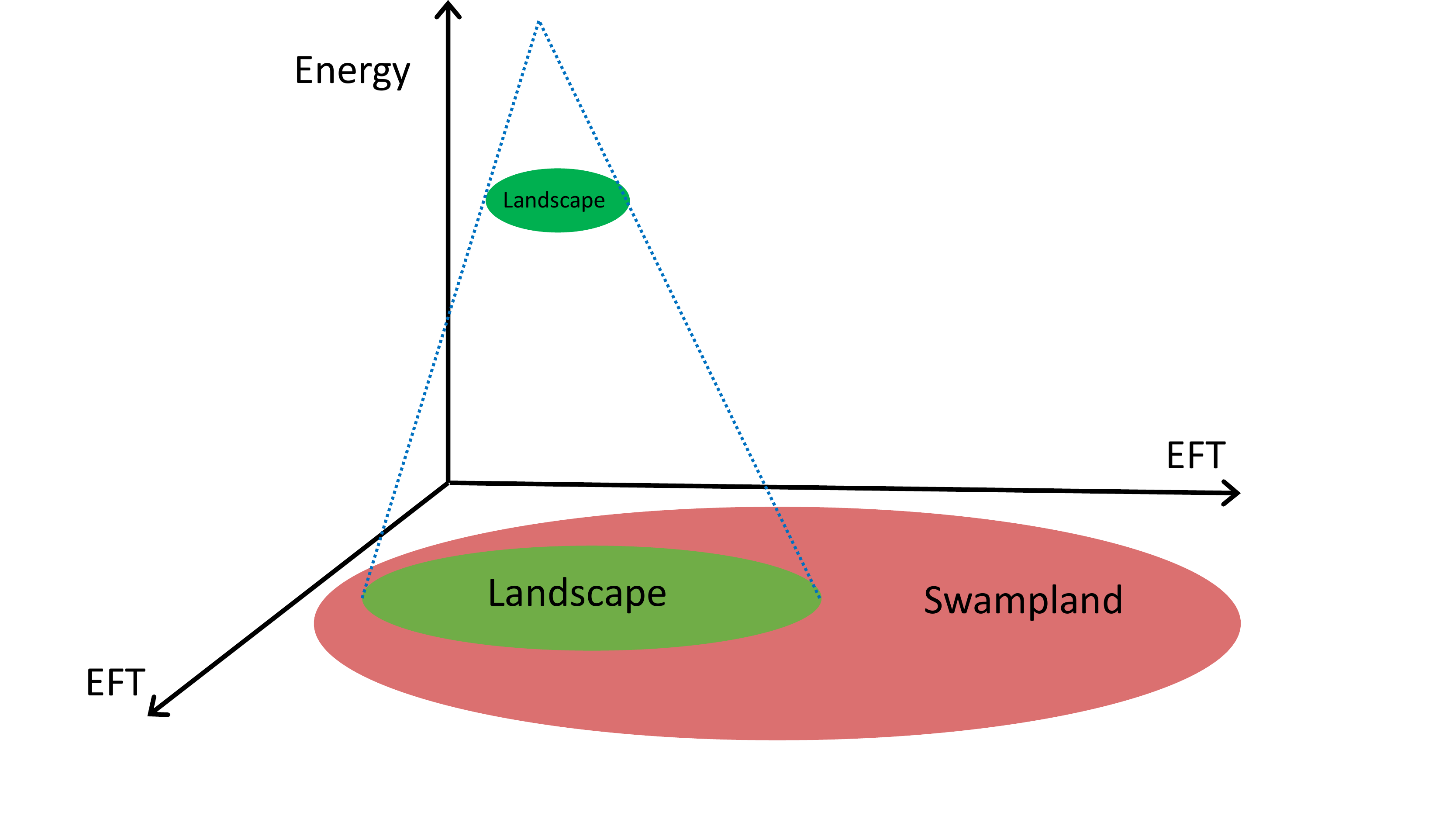} 
    \caption{\footnotesize{A picture illustrating the swampland vs the landscape in the space of EFTs.}}
    \label{fig:swampland}
\end{figure}

%\subsection{Swampland}
\end{enumerate}

\section{Conclusions and Outlook}
Clearly the field is in a very healthy condition, with concrete recent achievements and long term goals. In one way it has spread out in different subfields, from  formal aspects of field theory to explorations of deep questions on quantum gravity and information. There are interesting overlaps which may render fruitful results across the subfields. But the main questions such as the non-perturbative definition of string theory and  cosmological singularities, remain open. Plenty to look forward for future ICHEP conferences from this perspective. On the direction closer to experiments, we still hope for any discovery in present and future colliders as well as searches for axions, proton decay  and dark matter in order to get some guidance towards UV complete models. Also, there are expectations for future CMB experiments to give a hint on the value or limits on $r$, which could be informative about the scale of inflation and then of the fundamental theory behind. For the longer term,  exploring very high frequency gravitational waves \cite{Aggarwal:2020olq} looks very promising to test early universe physics beyond the Standard Model at energies beyond the reach of future colliders. 

\section*{Acknowledgements}
I would like to thank all my collaborators and colleagues for shaping my view on the subject. I particularly thank  Francesco  Muia, Veronica Pasquarella and Jorge Santos for comments on the manuscript. 
Research has been partially supported by STFC consolidated grants ST/P000681/1, ST/T000694/1.

\end{document}